\documentclass[useAMS,usenatbib]{mn2e}
\usepackage{epsfig,amsmath,amssymb,amsfonts,mathrsfs,latexsym,graphicx}
\include{journals}
\newcommand{\eg}{e.g.\ }
\newcommand{\ie}{i.e.\ }
\newcommand{\Msun}{{\rm M}_{\odot}}

\newcommand{\kms}{km~s$^{-1}$}

\newcommand{\OI}{O~{\sc i}}

\newcommand{\SiI}{Si~{\sc i}}
\newcommand{\SiII}{Si~{\sc ii}}

\newcommand{\TiII}{Ti~{\sc ii}}

\newcommand{\FeI}{Fe~{\sc i}}
\newcommand{\FeII}{Fe~{\sc ii}}
\newcommand{\FeIII}{Fe~{\sc iii}}
\newcommand{\CoI}{Co~{\sc i}}

\newcommand{\NiI}{Ni~{\sc i}}

\newcommand{\Feff}{$^{54}$Fe}
\newcommand{\Fefs}{$^{56}$Fe}
\newcommand{\Cofs}{$^{56}$Co}
\newcommand{\Nifs}{$^{56}$Ni}
\newcommand{\Nife}{$^{58}$Ni}

\newcommand{\KE}{$E_{\rm kin}$}
\newcommand{\Deltam}{$\Delta m_{15}(B)$}

\newcommand{\apj}{ApJ}
\newcommand{\apjl}{ApJL}
\newcommand{\apjs}{ApJS}
\newcommand{\aj}{AJ}
\newcommand{\aap}{A\&A}
\newcommand{\mnras}{MNRAS}
\newcommand{\nat}{Nat}
\newcommand{\pasp}{PASP}

\def\gsim{\mathrel{\rlap{\lower 4pt \hbox{\hskip 1pt $\sim$}}\raise 1pt \hbox {$>$}}}
\def\lsim{\mathrel{\rlap{\lower 4pt \hbox{\hskip 1pt $\sim$}}\raise 1pt \hbox {$<$}}}

\title[The nebular spectrum of SN\,2003hv]{The nebular spectrum of the 
type Ia supernova 2003hv:  evidence for a non-standard event}

\author[P.A. Mazzali et al.]{Paolo A. Mazzali$^{1,2}$\thanks{E-mail: 
mazzali@mpa-garching.mpg.de}, I. Maurer$^1$, 
M. Stritzinger$^{3,4}$, S. Taubenberger$^1$, S. Benetti$^2$, 
\newauthor
S. Hachinger$^1$ \\
\\
$^{1}$Max-Planck Institut f\"ur Astrophysik, Karl-Schwarzschildstr. 1, D-85748 
Garching, Germany \\
$^{2}$INAF-Osservatorio Astronomico, vicolo dell'Osservatorio, 5, I-35122 
Padova, Italy\\
$^{3}$Oskar Klein Centre \& Dept. of Astronomy, Stockholm University, Sweden \\
$^{4}$Dark Cosmology Centre, Niels Bohr Institute, University of Copenhagen, 
Juliane Maries Vej 30, 2100 Copenhagen \O, Denmark}

\begin{document}

\date{Accepted ... Received ...; in original form ...}

\pagerange{\pageref{firstpage}--\pageref{lastpage}} \pubyear{2011}

\maketitle

\label{firstpage}

\begin{abstract}

The optical and near-infrared late-time spectrum of the under-luminous Type Ia
supernova 2003hv is analysed with a code that computes nebular emission from a
supernova nebula. Synthetic spectra based on the classical explosion model W7
are unable to reproduce the large \FeIII/\FeII\ ratio and the low infrared flux
at $\sim 1$ year after explosion, although the optical spectrum of SN\,2003hv is
reproduced reasonably well for a supernova of luminosity intermediate between
normal and subluminous (SN\,1991bg-like) ones. A possible solution is that the
inner layers of the supernova ejecta ($v \lsim 8000$\,\kms) contain less mass
than predicted by classical explosion models like W7. If this inner region
contains $\sim 0.5 \Msun$ of material, as opposed to $\sim 0.9 \Msun$ in
Chandrasekhar-mass models developed within the Single Degenerate scenario, the
low density inhibits recombination, favouring the large \FeIII/\FeII\ ratio
observed in the optical, and decreases the flux in the \FeII\ lines which
dominate the IR spectrum.  The most likely scenario may be an explosion of a
sub-Chandrasekhar mass white dwarf. Alternatively, the violent/dynamical merger
of two white dwarfs with combined mass exceeding the Chandrasekhar limit also
shows a reduced inner density.

\end{abstract}


\section{Introduction}

Type Ia Supernovae (SNe\,Ia) are the most homogeneous type of stellar
explosions, and one of the most luminous. They are used to constrain the energy
content of the Universe and provided the first direct evidence of accelerated
expansion \citep{Reiss98,Perlmutter99}. SNe\,Ia are not standard candles per
se, but they can be standardised \citep[with a dispersion of $\sim 0.15$
mag;][]{Germany04,Folatelli10}, thanks to a rather tight relation between their
luminosity and the shape of their light curve. This was first suggested by
\citet{Phillips93}, who parametrised the light curves of SNe\,Ia based on a
single quantity, \Deltam, the decline of the $B$ band magnitude from maximum to
15 days later. 

SNe\,Ia come with a rather large range of luminosities. Their absolute
magnitudes range from $\sim -17$ for the dimmest events like SN\,1991bg
\citep{Filippenko92,Leibundgut93} to $\sim -20$ for the superluminous,
``super-Chandrasekhar" SNe \citep{Howell06,Hicken07,Scalzo10,Tauben11}, but the
bulk of spectroscopically normal SNe has a smaller spread, between $\sim -18.5$
and $\sim -19.5$. These are the majority of SNe Ia. At the luminous end of this
range confusion with the spectroscopically peculiar 1991T-like events may occur,
since these show similar decline rates and luminosities as some
spectroscopically normal SNe \citep[\eg SN\,1999aa,][]{Garavini04}. At the dim
end, however, peculiar events similar to SN\,1991bg, with \Deltam\,$\sim 2$\,mag
appear to be separated from even the dimmest of the normal SNe, such as
SNe\,1992A or 2004eo, which have \Deltam\, $\sim 1.4$\,mag. Only few SNe are
known that have intermediate decline rates \citep[\eg][]{Tauben08}. A
classically known example is SN\,1986G  
\citep[\Deltam\,$\sim 1.8$\,mag,][]{Phillips87}, which however shows many  
properties of the SN\,1991bg group, in particular the presence of a strong  
\TiII\ absorption trough between 4000 and 4500\,\AA\ at maximum light and the 
evolution of the \SiII\ absorption velocity \citep{Benetti05}. 

Since 1991bg-like SNe\,Ia are almost exclusively observed in early-type galaxies
\citep{Hamuy00} the question arises whether they may have entirely different
progenitors. For the bulk of SNe\,Ia a carbon-oxygen (CO) white dwarf accreting
hydrogen in a binary system has traditionally been the scenario of choice,
because it is thought to be able to give rise to homogeneous explosions. The
observed range of SN luminosities can be attributed to the synthesis of
different amounts of \Nifs\ in the explosion
\citep{Arnett82,Branch92,Riess96,Cappellaro97,Contardo2000,Zorro,Stritzinger06}.
The most promising way to tune this parameter is given by the explosion scenario
known as Delayed Detonation \citep{Khokhlov91}, although the details of the
transition from subsonic to supersonic burning are not clear. A recently revived
alternative scenario is that some fraction of the SNe\,Ia come from the merging
of two white dwarfs whose combined mass exceeds the Chandrasekhar mass
\citep{IbenTutukov84,Webbink84,Pakmor10}. Another scenario is the explosion of 
a white dwarf with a mass below the Chandrasekhar limit. This is thought to
occur if the white dwarf accretes helium from a companion. the helium shell
which builds up is very unstable and can detonate, triggering the explosion of
the entire star \citep{WoosleyWeaver94,LivneArnett95}. This scenario was not
favoured because it was reported to lead to inconsistent SN spectra  
\citep{Nugent97}. However, \citet{Sim10} present sub-Chandrasekhar SN\,Ia models
where the overlying He layer is omitted and suggest that these models reproduce
the observed range of light curve properties of SNe\,Ia (from intermediate to
dim), and that spectra obtained from their models also resemble observed
spectra.  

Still, most SNe\,Ia, excluding perhaps the least luminous, 1991bg-like ones,
seem to be consistent with the Chandrasekhar mass \citep{Zorro}. 

A powerful method of investigating the properties of SN ejecta is late-time
spectroscopy, which probes the inner layers of the SN. \citet{Mazzali10} showed
that the simultaneous availability of optical and infrared (IR) late-time
spectra makes it possible to account for all the most abundant elements produced
in a SN, and to infer their abundances. Not many late-time IR spectra are
available for SNe\,Ia. One of the best examples is SN\,2003hv.
\citet{Motohara06} published IR spectra which \citet{Leloudas09} rescaled to
match IR photometry taken at an epoch similar to that of a mid-IR spectrum
\citep{Gerardy07}.  

This is not the only thing that makes SN\,2003hv interesting. The SN has a
decline rate \Deltam\,$= 1.61$\,mag \citep{Leloudas09}. A SN\,Ia with such a
decline rate could be the missing link between spectroscopically normal SNe\,Ia
and underluminous SNe of the 1991bg class. SN\,2003hv may be the first
well-observed SN with this decline rate. \citet{Leloudas09} presented a very
good data set covering the post-maximum phase. 

Additionally, SN\,2003hv was noted for the peculiar shifts of several nebular IR
emission lines, some of which also seem to show flat tops
\citep{Gerardy07,Motohara06,Maeda10a}. \citet{Leloudas09} find that similar
displacements are observed in the optical lines of [\FeII]. This was interpreted
as possibly the indication of a high-density neutron-capture region, where
stable \Nife\ and \Feff\ rather than radioactive \Nifs\ are produced. This
region would peculiarly be displaced from the centre of the explosion
\citep{Maeda10a}. This finding led to further work in this direction
\citep{Maeda10b}, which is however not the subject of the present paper. 


\citet[][Fig.7]{Leloudas09} showed a synthetic optical/near-IR spectrum based on
the classical explosion model W7 \citep{Nomoto84}. This is a Chandrasekhar-mass
model, with kinetic energy \KE $\sim 1.3 \times 10^{51}$ erg, which was
developed to account for the bulk properties of SNe\,Ia \citep{Branch85} and has
been used as a standard reference to describe this class of SNe. 
While the overall reproduction of the spectrum is good, they encountered a few
problems, which they mentioned and discussed  \citep{Leloudas09}. Most
evidently, the near-IR flux ($\lambda > 7500$\,\AA) is overestimated by the
model by a factor of a few. Additionally, the ratio of the two strongest optical
features, the Fe-dominated lines near 4700 and 5200\,\AA, is not reproduced.
While the bluer feature is dominated by [\FeIII] lines, the redder one is
predominantly due to [\FeII] transitions. The bluer feature is well reproduced
in \citet{Leloudas09}, while the redder one is overestimated by about a factor
of 2. Since most of the near-IR lines whose flux is also overestimated are
themselves [\FeII] transitions this seems to indicate that the ionization
balance in the SN\,2003hv nebula is different from that predicted  by W7.  Among
other possibilities, \citet{Leloudas09} suggest that the lack of near-IR flux
might be attributed to a sudden onset of an infrared catastrophe in the densest
region of clumpy ejecta. The IR catastrophe (IRC) is a sudden shift of the
nebular emission to mid-IR lines, which occurs when the temperature of the SN
nebula becomes sufficiently low \citep{Axelrod80}.  \citet{Leloudas09} argue
that an IRC may have occurred at earlier times, and that this may be possible if
the ejecta are clumpy. They also find that positrons are likely to be fully
trapped.

Given the special role of SN\,2003hv as a possible link between normal and
subluminous SNe\,Ia, and the significance of the late-time data available, we 
have tried to reproduce these data with our nebular code. 
This paper is organised as follows: 
in Section 2 we briefly describe our nebular code;
in Section 3 we show and discuss synthetic spectra obtained adopting different
scenarios; 
in Section 4 we discuss the time evolution of the nebular emission;
in Section 5 our results are recapped and discussed. 
Finally, Section 6 concludes the paper.

\section{Method}

\citet{Leloudas09} present a combined optical, near-IR and mid-IR spectrum of
SN\,2003hv in the nebular phase. The optical spectrum was obtained at the
ESO-VLT 320 days after $B$-band maximum. The near-IR spectrum was obtained at
Subaru 394 days after $B$-band maximum \citep{Motohara06}, while the mid-IR
spectrum was obtained from {\em Spitzer} 358 days after $B$-band maximum
\citep{Gerardy07}. Since the optical and IR data were not obtained
simultaneously, \citet{Leloudas09} rescaled the optical spectrum and the near-IR
spectrum to match the photometry of day 355 after maximum, the epoch of the
mid-IR observations. Given the relatively short time elapsed between the optical
and near-IR observations, it can be hoped that the spectrum did not evolve much
in shape. We use the data presented in Fig. 7 of \citet{Leloudas09}. 

At these late times, the spectrum of SN\,2003hv is dominated by  nebular
emission. The gas is optically thin to optical and IR radiation, and it is
heated by the deposition of the $\gamma$ rays and positrons from the decay of
\Cofs\ to \Fefs\ (\Nifs\ has almost completely decayed to \Cofs). Collisional
heating is balanced by cooling via emission, and the energy is released in a
number of line transitions, most of which are forbidden. 

We use a non-local thermodynamic equilibrium (NLTE) nebular code which was
developed by \citet{Mazzali01} and is based on ideas of \citet{Axelrod80} and
\citet{RLL92}. Although a three-dimensional version is available
\citep{Maureretal10a} the code was used in its one-dimensional version here. Our
(one-dimensional) models consist of about 20 radial shells of varying density
and composition. The emission and deposition of $\gamma$ rays and positrons is
computed with a Montecarlo scheme, following \citet{Cappellaro97} and
\citet{Lucy05}, using effective opacities for $\gamma$ rays and positrons.

After energy deposition, an iteration process is initiated. Starting with an
initial guess for the ionisation state and the temperature of the gas,
non-thermal ionisation and excitation rates are derived using the Bethe and the
optical approximation respectively \citep{Axelrod80,Rozsnyai80,MaurerMazzali10},
which have been shown to be accurate to at least 20\% for H and He
\citep{Maureretal10b}. For other elements the accuracy of these approximations
is difficult to estimate, since there is a strong dependence on poorly known
atomic data. Since our nebular code neglects radiation transport, there is no
photo-ionisation. The collisional ionisation rates are balanced with the
radiative and the di-electronic recombination rates taken from
\citet{AldrPeq73}. Charge exchange reactions are not considered. After obtaining
the new ionisation state and consequently a new electron density, the level
population of each ion is derived by solving  a matrix containing collisional
(de-)excitation and radiative de-excitation rates. The radiative rates are
reduced according to the Sobolev optical depths of the corresponding lines.
Knowing all the level populations, a radiation field can be generated and can be
compared to the total deposited luminosity. The electron temperature (and
consequently the electron density) is then adjusted in each iteration step,
until the total luminosity of the radiation field has converged to the deposited
luminosity.

The code can treat the neutral and singly ionized states of H, He, C, N, O, Ne,
Na, Mg, Si, S, Ar \& Ca and the singly and doubly ionized states of Fe, Co \&
Ni. For light and intermediate-mass elements typically the lowest 10 $-$ 20
levels are taken into account in the excitation calculation. For iron-group
elements roughly the 100 lowest energy levels of each ion are considered.
Collisional data are available for several hundred transitions of H, He, C, N,
O, Ne, Na, Mg, Si, S, Ar, Ca, Fe, Co \& Ni. Most of the collisional data are
taken from \citet{Axelrod80} or from online databases (e.g.
TIPTOP\footnote{http://cdsweb.u-strasbg.fr/OP.htx}, A.
Pradhan\footnote{http://www.astronomy.ohio-state.edu/$\sim$pradhan/}) and are
updated frequently. Collisional data are available for the most important
forbidden lines, but those for Co are unfortunately very poor. If not available,
the collision strength of forbidden lines is crudely set to
0.01$g_\mathrm{l}g_\mathrm{u}$ \citep{Axelrod80}, where $g_\mathrm{l}$ and
$g_\mathrm{u}$ are the statistical weights of the lower and the upper atomic
level respectively. Unknown collisional data for allowed transitions are
obtained in the \citet{VanRegemorter62} approximation. 

Although missing collisional data pose a problem for all kinds of nebular
calculations, the largest uncertainties in our calculations are caused by
neglecting photo-ionization. It is clear that the neutral species of the
iron-group elements are photo-ionized almost completely in SNe Ia at the epochs
of interest (200 - 400 days after the explosion). To simulate this effect, \FeI,
\CoI\ and \NiI\ are not included in our calculation. The ratio of \FeIII\ to
\FeII\ is determined by the non-thermal electron ionisation rates, which is a
good approximation. For light and intermediate-mass elements the problem is more
serious and it is not clear if we can produce the ionisation state of those
elements correctly. However, since our synthetic spectra are dominated by \FeII\
and \FeIII\ emission lines, we expect them to be reliable within the
uncertainties of the atomic data.  

The code has been used to investigate the morphology and composition of the
spectra of various SNe \citep[\eg][]{Mazzali05,Mazzali07b}.  A 1-zone version
makes it possible to derive very quickly the basic properties of the emitting
nebula (mass, energy, composition) using the maximum observed emission line
velocity as an outer boundary. In a more sophisticated stratified version the
density varies as a function of velocity (which is equivalent to radius in the
homologously expanding SN nebula) and so do the abundances. These can be
modified in order to reproduce the observations, yielding a rather accurate
description of the structure of the SN ejecta. The density structure can be
taken from a model calculation \citep[\eg W7,][]{Tanaka11}, but it can also be
derived from the emitted luminosity and the line shapes. 


For all calculations we adopt for SN\,2003hv a distance modulus of 31.58 mag
\citep{Leloudas09} and a reddening $E(B-V) = 0.016$\,mag \citep{Schlegel98}.
Reddening within the host galaxy is negligible \citep{Leloudas09}.

\begin{figure}
\includegraphics[width=89mm]{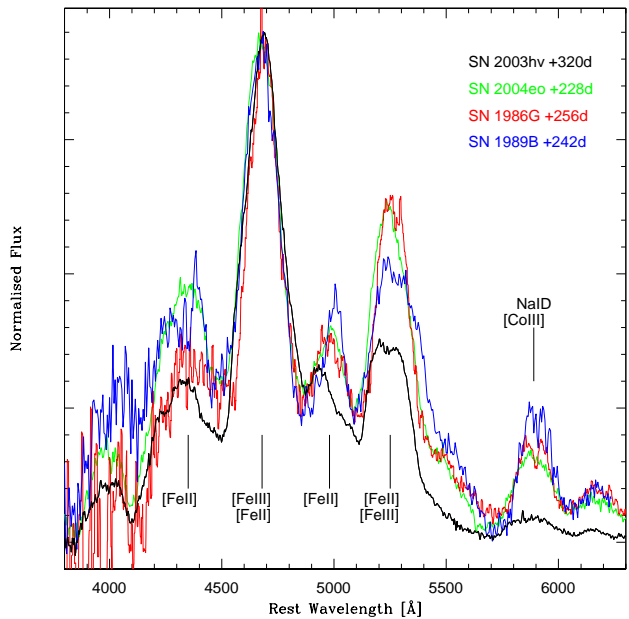}
\caption{Late-time optical spectra of 4 rapidly declining SNe\,Ia: 
SN\,2003hv (\Deltam\,$ = 1.61$\,mag, spectroscopically normal) - black;
SN\,2004eo (\Deltam\,$ = 1.47$\,mag, spectroscopically normal) - green;
SN\,1989B (\Deltam\,$ = 1.28$\,mag, spectroscopically normal) - blue;
SN\,1986G (\Deltam\,$ = 1.73$\,mag, spectroscopically peculiar) - red.
The spectra have been normalised at the peak of the 4700\,\AA\ emission.}
\label{spectra}
\end{figure}

\section{Models}

The late-time optical spectra of SNe\,Ia are dominated by strong emission in
forbidden lines at $\sim 4700$ and 5200\,\AA. The features are broadened
because they are blends of many lines, so that their width is larger than the
real velocity of the emitting gas. The bluer feature is dominated by [\FeIII]
lines ($\lambda\lambda 4607$, 4658, 4667, 4701, 4734, 4755, 4881), with some
[\FeII] lines ($\lambda\lambda 4814$, 4874, 4890) also contributing, while the
redder one has comparable contributions of [\FeII] ($\lambda\lambda 5111$, 5159,
5222, 5262) and [\FeIII] lines ($\lambda\lambda 5270$, 5412). A correlation
exists between the width of the features and the luminosity of the SN as
described by the light-curve decline paramater \Deltam\ \citep{Mazzali98}. This
is because a more luminous SN burns more of the progenitor white dwarf into
\Nifs, which is observed at late times as Fe emission. The availability of the
near-IR spectrum makes it possible to study all elements that are supposed to be
abundantly synthesised in a thermonuclear explosion.  

To begin with, we compare the optical spectrum of SN\,2003hv to those of other
SNe\,Ia with similar decline rates. Figure 1 shows the blue part of the nebular
spectra of 4 SNe\,Ia: SN\,2003hv (\Deltam\,$ = 1.61$\,mag), 
SN\,1986G \citep[][\Deltam\,$ = 1.73$\,mag]{Phillips87},  
SN\,2004eo \citep[][\Deltam\,$ = 1.47$\,mag]{Pasto07},  
and SN\,1989B \citep[][\Deltam\,$ = 1.28$\,mag]{Wells94}.  
With the exception of SN\,2003hv all spectra are from the Asiago database. 
They have all been corrected for reddening, and are plotted after normalizing at
the peak of the 4700\,\AA\ line. SN\,1986G has the narrowest emission lines, as
may be expected given that this SN has the most rapidly declining light curve.
The other 3 SNe have all rather similar line width. SN\,2003hv stands out for
the much higher ratio of the 4700\,\AA\ emission feature with respect to the 
5200\,\AA\ one.

We begin with modelling the optical spectrum alone. For this purpose we use the
optical spectrum calibrated to the original photometry of the epoch when it was
actually observed, 320 days after maximum, thus avoiding any unforeseen
evolution between then and the epoch of the near- or mid-IR spectrum. We discuss
here various models.

\subsection{One-zone models} 

We use the one-zone version of the code in order to determine the velocity width
of the emitting nebula, assuming that it is spherically symmetric. The
emission lines of SN\,2003hv can be reproduced for a width of the emitting
region of $8000$\,\kms. We also measure the FWHM of the 4700\AA\ line, finding a
value of $13500\pm450$\,\kms. Both of these values are consistent with Fig.2 of
\citet{Mazzali98}, indicating that SN\,2003hv is located at the edge of the
distribution of normal SNe\,Ia, close to SNe 1986G and 1993L. 
The value of \Deltam\ that would be obtained from the fit
shown in Figure 2 of \citet{Mazzali98} is $\sim 1.4$\,mag. 
The photometric value \citep[\Deltam\ $= 1.61\pm0.02$\,mag;][]{Leloudas09} is 
different but not inconsistent, given the large dispersion of the relation
shown in \citet{Mazzali98}. 

However, SN\,2003hv stands out because it shows a much larger ratio of the
bluer, \FeIII-dominated line near 4700\,\AA\ over the redder, \FeII-dominated
one near 5200\,\AA\ (see Fig.1). Reddening towards SN\,2003hv is small. 
SNe\,1989B and 1986G are both highly reddened \citep[$E(B-V) = 0.37$ and
0.60\,mag;][respectively]{Wells94, Phillips99}, but even when this is corrected
for a clear difference remains. The same is true if we compare SN\,2003hv with
SN\,2004eo, which has a lower reddening \citep[$E(B-V) =0.11$\,mag,][]{Pasto07}
and similar line width. 

A reasonable 1-zone model for the optical spectrum of SN\,2003hv is obtained for
a \Nifs\ mass of $0.42 \Msun$ and is shown in Figure 2. This is actually rather
a large value for the SN decline rate: for SN\,2004eo, with \Deltam$ =
1.47$\,mag, \citet{Mazzali08} found M(\Nifs) $= 0.35 \Msun$. However, this value
is consistent with the estimate of \citet{Leloudas09} based on the luminosity at
peak. The main reason for the large \Nifs\ mass is indeed the need to fit the
strong \FeIII\ emission. The large \FeIII/\FeII\ ratio ($\sim 1 : 1$) which
characterizes the model is obtained assuming that no stable \Feff\ is present in
the centre. This isotope acts only as a coolant and its effect is to reduce the
ionization of Fe. Not introducing stable Fe keeps the ionization ratio high. Co
and Ni have a similar ionization state as Fe. Intermediate-mass elements (IME)
are more highly ionized. The mass used in the model within 8000\,\kms\ is only
$0.46 \Msun$, which is less than in 1-zone models for the other SNe (e.g.,
SN\,2004eo requires $\sim 0.55 \Msun$ within 7300\,\kms) and significantly less
than the prediction of W7 \citep[$0.63 \Msun$,][]{Iwamoto99}. So, in the 1-zone
model, SN\,2003hv has a small mass at low velocities, and most of it seems to
have been burned to \Nifs.

Although the 1-zone model does a reasonable job of reproducing the optical
spectrum, it can be seen to fail already in the reddest part: the [\FeII]
emission lines redwards of $\sim 7500$\,\AA\ are overestimated. This becomes
even worse in the near-IR, as is shown in Figure 2. 

There are therefore several different lines of evidence indicating that
SN\,2003hv may not conform with the norm of SNe\,Ia. This can be further studied
using stratified models. 

\begin{figure*}
\includegraphics[width=160mm]{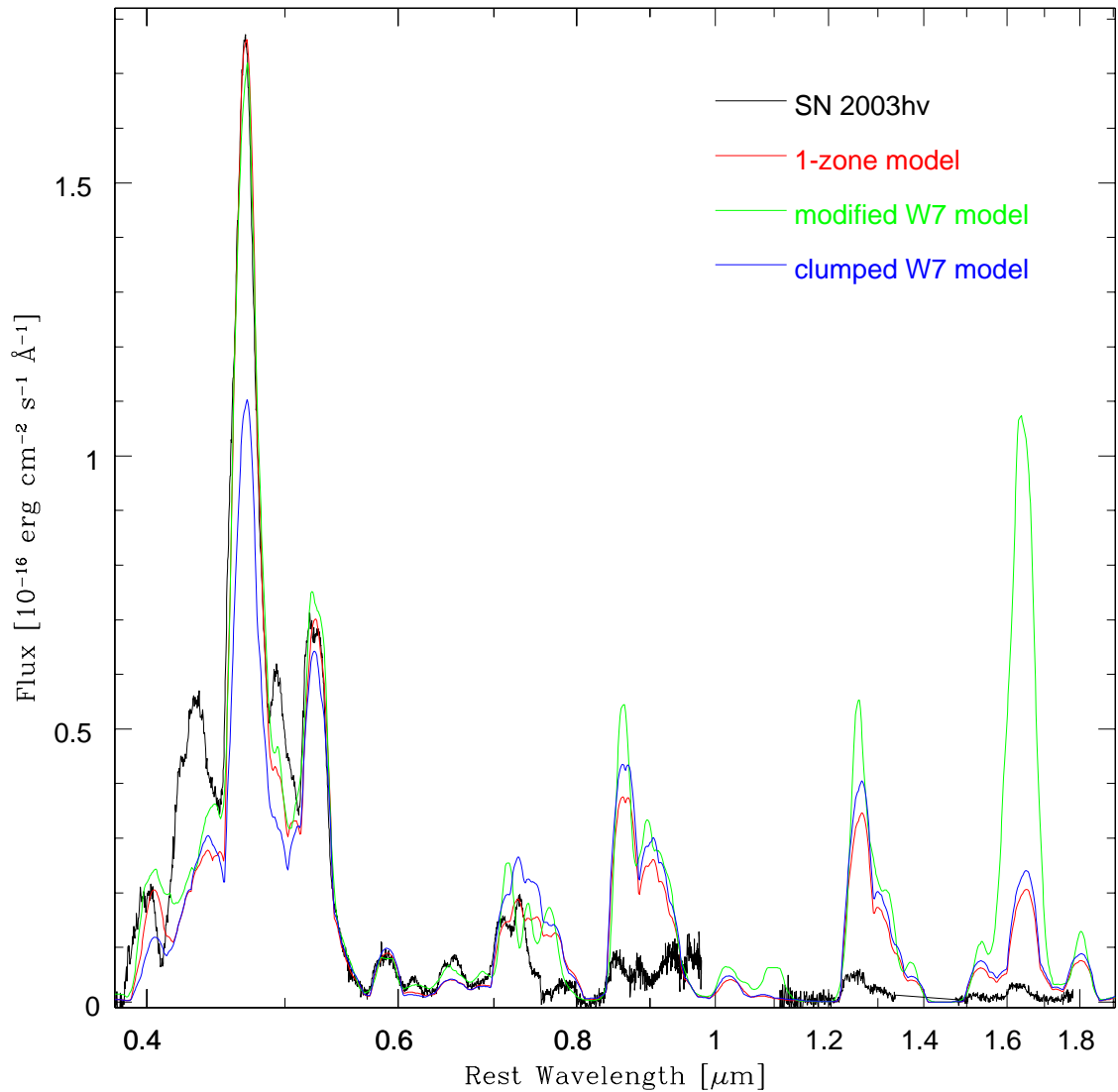}
\caption{The late-time optical/near-IR spectrum of SN\,2003hv compared to 
various synthetic spectra:
a 1-zone model (red);
a stratified model based on the W7 density distribution but with
modified abundances (green);
a stratified model also based on W7 but with the additional inclusion of
clumping in density (blue). }
\label{modelsW7}
\end{figure*}

\subsection{Stratified models: W7} 

In the stratified version, our nebular code can be used to test specific
explosion models, or, as we do here, to construct a best-fitting
density-abundance distribution by fitting observations. \citet{Zorro} fitted the
late-time spectra of 23 SNe\,Ia of different luminosities using the density
distribution of W7. They changed the \Nifs\ content by modifying the outer
extent of this species, as guided by the width of the observed nebular lines of
Fe. They found that not only the mass and distribution of \Nifs\ correlates with
the luminosity: evidence that the distribution of IME such as silicon and
sulphur is complementary to that of nuclear statistical equilibrium (NSE)
material suggests that most of the material which is not burned to Fe-group
elements must have been burned to IME. This supports the notion of a common mass
for all SNe\,Ia, or at least for most of them, excluding perhaps 1991bg-like
events \citep{Hachinger09}. 

Here we test whether the W7 density distribution can be compatible with
SN\,2003hv. The W7 model \citep{Nomoto84} was created to fit the basic
properties of average SNe\,Ia. The underlying assumption is that SNe\,Ia result
from the deflagration of Chandrasekhar-mass C-O white dwarfs. We now know that
this is at best a secondary channel to the production of SNe\,Ia \citep{Sahu08}.
However, the density-velocity distribution of W7 is representative of typical
SN\,Ia explosions: delayed detonation models are indeed very similar
\citep{Iwamoto99}.

Using the W7 density structure, and modifying only the abundances as in
\citet{Zorro}, it is possible to get a reasonable fit of the optical spectrum of
SN\,2003hv (Figure 2). This model has M(\Nifs)\,$ = 0.42 \Msun$, similar to the
1-zone model. \Nifs\ is located between 2000 and 10000\,\kms, with abundances
peaking at $\sim 70$\,\% at 5000-7000\,\kms\  (Figure 3). Stable \Feff\ is
located only in the innermost region, below 4000\,\kms: the total mass of \Feff\
is $0.09 \Msun$, as in W7 \citep{Iwamoto99}.  IME must already be present as
deep as $\sim 4000$\,\kms\ and become dominant above $\sim 8000$\,\kms,
otherwise the mass of \Nifs\ would be too large and the Fe lines too broad.
Regions above 10000\,\kms\ are not sensitive to the nebular treatment as their
density is too low. 

The synthetic spectrum basically reproduces the results of \citet{Leloudas09},
showing that the two codes give consistent results. While the blue part of the
spectrum reproduces SN\,2003hv, the emission lines in the red part of the
optical spectrum as well as in the near-IR are obviously too strong (Figure 2).
Most of the lines in those regions are due to [\FeII]. The strongest features in
the bluer part of the optical spectrum, on the other hand, are either dominated
by [\FeIII] (the line near 4700\,\AA) or have a non-negligible [\FeIII]
contribution (the line near 5200\,\AA). This suggests that the ionization degree
of Fe is too low in the model. This cannot be improved as long as the high
central density that characterizes W7 is kept. Another reason for the low
ionization is that in the inner part of the ejecta stable Fe-group isotopes
dominate. The role of these isotopes is only that of a coolant. Replacing stable
Fe-group isotopes with \Nifs\ would increase the ionization because of the added
heating, but it would also lead to an increase of the luminosity. Replacing them
with IME would further increase the [\FeII]-[\SiI] line near $1.6 \mu$m, which
is already too strong (Figure 2), and replacing them with oxygen would cause an
emission line due to [\OI] 6300, 6363\AA\ \citep{Kozma05}, which is not
observed.  In any case, none of these changes would be justified by explosion
physics, since some neutron-rich material should be synthesised at the high
central density of a Chandrasekhar-mass white dwarf.

\begin{figure}
\includegraphics[angle=-90,width=90mm]{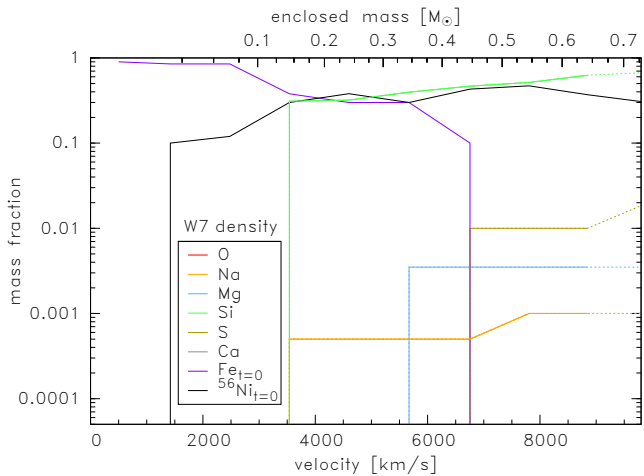}
\caption{Abundances for the model based on the W7 density distribution. }
\label{abuW7}
\end{figure}

\subsection{Stratified models: clumped W7 } 

\citet{Leloudas09} suggest that if the inner ejecta were clumped the lack of
near-IR radiation may be explained because cooling would be less efficient than
in the absence of clumping.  Our code allows the simulation of clumping in
density: if we define a filling factor $f < 1$, the gas is assumed to be
concentrated in clumps which fill a fraction $f$ of the volume.  We tested a
density clumping scenario with a moderate filling factor of 0.5, which does not
affect the $\gamma$-ray deposition. The result of this is that the higher
density in the clumps actually favours recombination, and further decreases the
\FeIII/\FeII\ ratio. A model with clumping is shown in Figure 2.  The synthetic
spectrum is still characterised by a high ratio of near-IR to optical flux.
Therefore, moderate clumping in density is unlikely to explain the observations
of SN\,2003hv.

\begin{figure*}
\includegraphics[width=160mm]{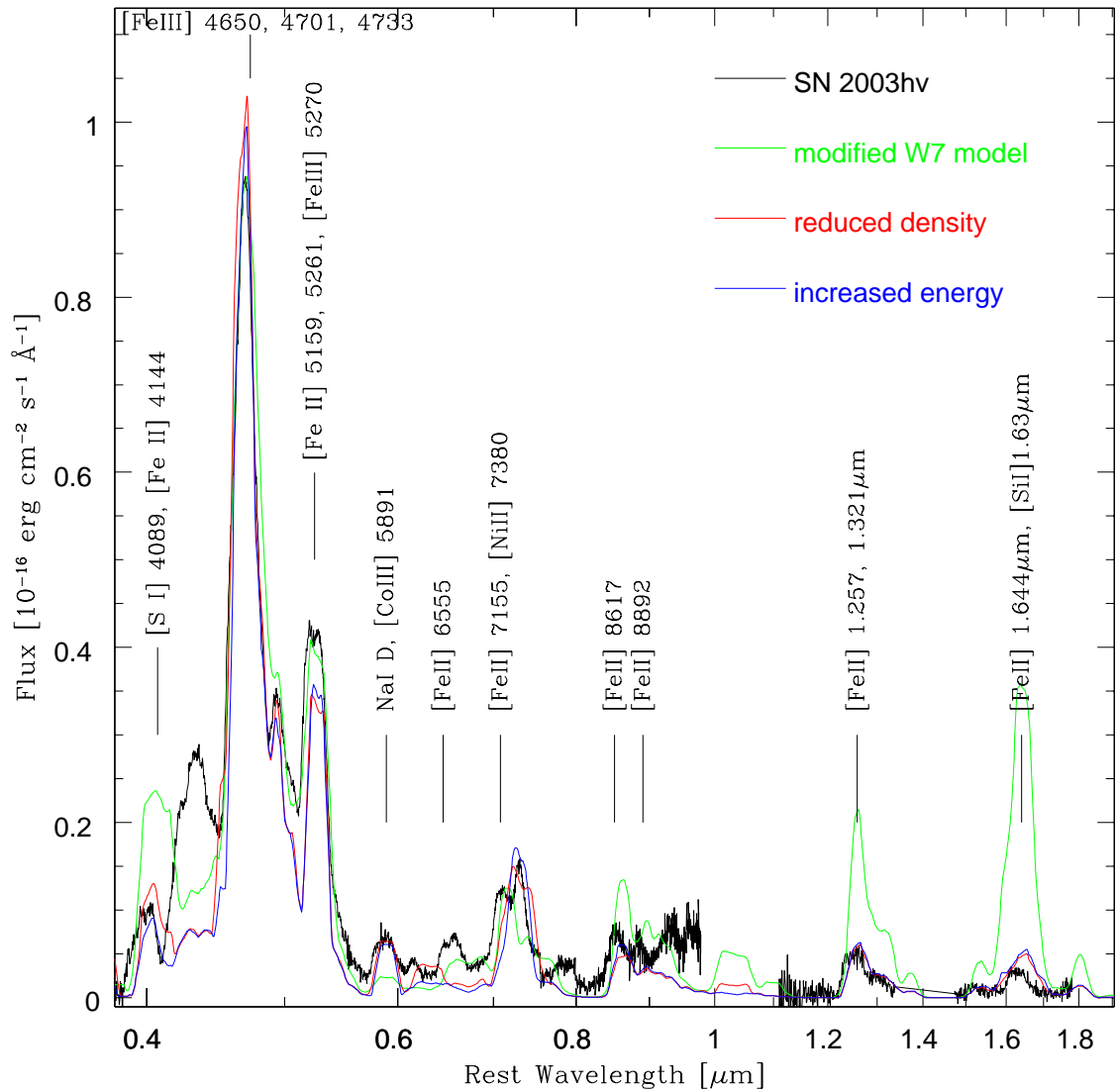}
\caption{The late-time optical/near-IR spectrum of SN\,2003hv, scaled to the 
photometry of day 355 after $B$ maximum (black) compared to various synthetic 
spectra.
Green: a stratified model based on the W7 density distribution but with
modified abundances;
red: a stratified model with a reduced density in the inner layers;
blue: a stratified model with reduced density in the inner layers but an 
increased density in the outer layers to preserve the Chandrasekhar mass. }
\label{models}
\end{figure*}

\subsection{Stratified models: a modified density}

Since there seems to be no solution for the spectrum of SN\,2003hv using a W7
density, we next investigate whether different density structures might lead to
better results. The advantage of modelling nebular-phase spectra of SNe\,Ia is
that photoionization can be, to a first approximation, neglected in the iron
core (see Section 2). Combined with the homologous expansion of the SN ejecta,
this makes nebular spectroscopy a very powerful tool.  If the properties
(density and abundances) of a certain velocity shell are modified while the rest
of the nebula is left unaffected it is possible to derive the properties of that
specific velocity shell. Thus we can test in a rather uncomplicated way whether
modifications in the density structure can lead to an ionization ratio which
favours \FeIII\ more than W7 does. 

Qualitatively, in order to increase the ionization of the gas the recombination
rate should be reduced, which could be achieved with lower densities.
Alternatively, the ionization rate should be increased, which can be obtained by
reducing the ratio of stable versus radioactive NSE elements.  This suggests
that the density in the regions where the lines are emitted may be smaller than
in W7 \citep[see also][]{Maeda10a}, and that this may best be achieved by
reducing the content of stable NSE material, as already found in the one-zone
model (Section 3.1). Therefore we arbitrarily decreased the W7 density in the
inner part of the ejecta, where the emission lines are produced.  This was
achieved  mostly by removing stable NSE material. The outer layers, which cannot
be probed in the nebular phase because their density is too low, were left as in
W7.  At every step, we also modified the abundances in order to obtain as good a
match to the data as possible, both in the optical and, in particular, in the
near-IR. We modelled the optical/near-IR spectrum, now rescaled to a common
epoch of day 355 after $B$ maximum following \citet{Leloudas09}.  The modified
model has only $\approx 0.5 \Msun$ of material with $v < 10000$\,\kms.  In W7,
this is $\approx 0.9 \Msun$. 

The resulting synthetic spectrum is shown in Figures 4 and 5 (red line). As
expected, we indeed find that when a reduced density is used the ionization
degree is increased with respect to the W7-based model (Fig.4, green line). This
leads to a reduced flux in the red and the near-IR, which is the main result we
tried to achieve. At the same time, the increased ionization causes the [\FeIII]
4700\,\AA\ line to become somewhat stronger with respect to the 5200\,\AA\ line,
where both [\FeIII] and [\FeII] lines contribute. The ratio of these two Fe
lines is not perfectly reproduced, as we tried to obtain a synthetic spectrum
that reproduced the observed one as well as possible over a wide wavelength
range. A major deficiency is the line near 4300\,\AA\, which is much too weak. A
synthetic [\FeII] line at 4416\AA\ is present, but it is not very strong.
[\FeII] 6555\,\AA\ is also too weak. Increasing the strength of those lines
would lead to the overestimate of other [\FeII] lines, as in \citet{Leloudas09}.
The emission line near 1.6\,$\mu$m does not show a flat top: this would require a
three-dimensional study \citep[see][]{Maeda10a} since this and other features
characterized by a flat top are peculiarly blueshifted in the data
\citep{Motohara06}.  In our model $\gamma$ rays deposit in the central zone, and
flat tops are not expected. Figure 5 shows a blow-up of the red/near-IR part of
the spectrum.  


The various density profiles tested in the paper are shown in Figure 6.   The
\Nifs\ mass included in the reduced-density model is only $0.18 \Msun$, and the
mass of stable NSE elements (mostly \Feff) is $0.05 \Msun$, in both cases much
less than in W7 \citep[which has a \Nifs\ mass of $\sim 0.63 \Msun$ and a stable
NSE mass of $\sim 0.2 \Msun$,][]{Iwamoto99}. Some reduction of the \Nifs\ mass
is always seen when going from 1-zone models to stratified ones, and it results
from the increased energy deposition in the densest inner layers of a stratified
model \citep[\eg][]{Mazzali07b}. In the case of SN\,2003hv, however, this
reduction is dramatic. It is explained by the significantly reduced near-IR
luminosity in the new model. There is now a large discrepancy with the estimate
from the maximum of the light curve, a result also obtained by
\citet{Leloudas09} when comparing the peak and the tail of the light curve. 
Figure 7 shows the abundance distribution for this model, and Figure 8 the
ionization of some key elements, compared to the model based on the W7 density
distribution.

\begin{figure}
\includegraphics[width=88mm]{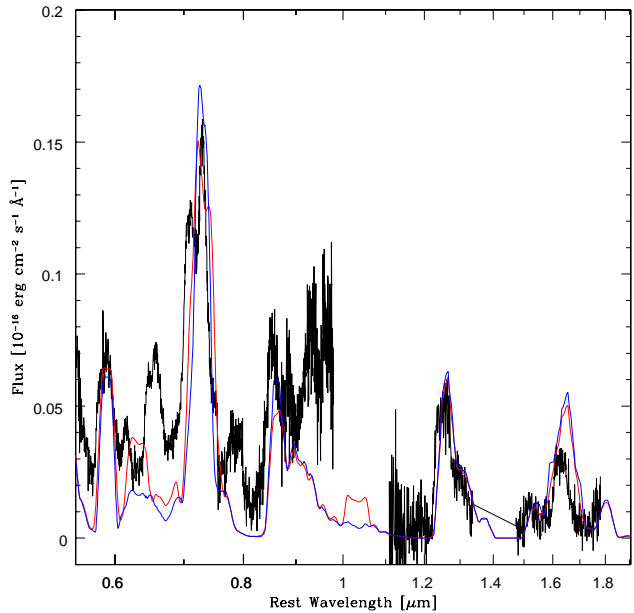}
\caption{Same as above, but focussing on the low-level red and near-IR part of
the spectrum.
The two models shown are 
the stratified model with a reduced density in the inner layers (red);
the stratified model with reduced density in the inner layers but an 
increased density in the outer layers to preserve the Chandrasekhar mass (blue). }
\label{models_low}
\end{figure}

\begin{figure}
\includegraphics[width=88mm]{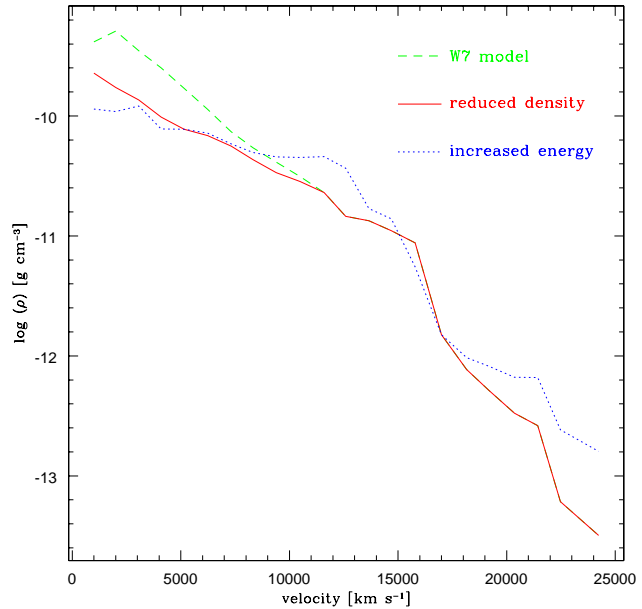}
\caption{The density structures used for the calculations:
the W7 density distribution (green);
a modified model with a reduced density in the inner layers (red);
a model with reduced density in the inner layers but an increased
density in the outer layers to preserve the Chandrasekhar mass (blue). }
\label{density}
\end{figure}

\begin{figure}
\includegraphics[angle=-90,width=90mm]{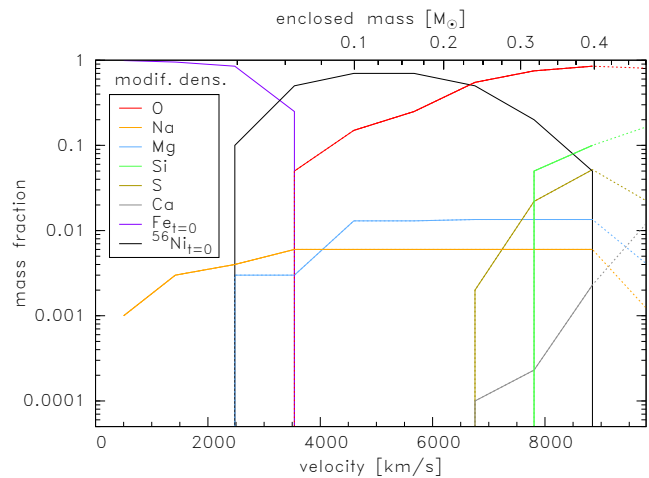}
\caption{Abundances for the model with the modified density distribution. We
show the part of the ejecta which can be probed by nebular spectroscopy. }
\label{abu}
\end{figure}

Finally, we computed a spectrum for a model where the mass which is missing at 
low velocities, compared to the W7 model, is located at high velocities. The
mass which has been removed from the densest inner layers is added at higher
velocity in this model (Figure 6) so that the width of the emission lines is not
increased. This model produces a similar nebular spectrum as the one with
reduced mass (Figures 4 and 5). Nebular models cannot explore the properties of
the outer layers. This requires modelling the early-time spectra
\citep[\eg][]{Hachinger09}. Only then may a complete model for the SN be
defined.

\begin{figure}
\includegraphics[width=88mm]{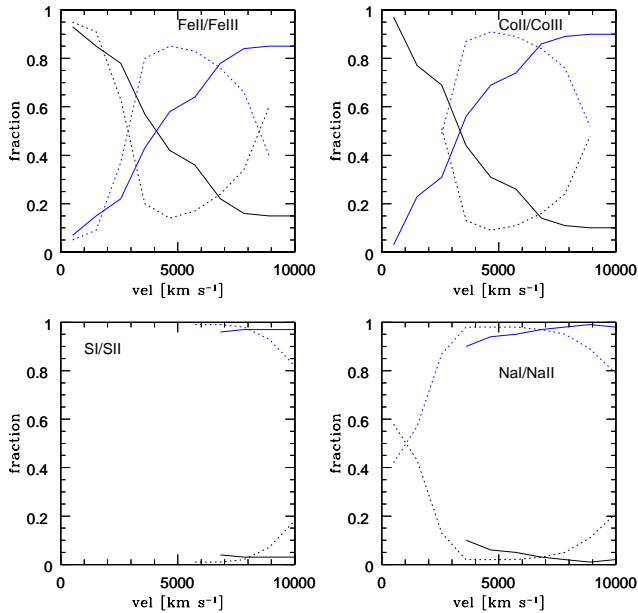}
\caption{The ionization structures for the W7 density distribution 
(dotted lines)
and the modified model with a reduced density in the inner layers (fully drawn
lines). In the various panels, blue lines represent the more highly ionised of
the two species treated. Truncation of the lines reflects the lack of the
corresponding element in certain velocity shells. }
\label{ionization}
\end{figure}

\section{Evolution of the optical/near-IR spectrum}

We have shown that the optical/near-IR spectrum of the underluminous but
spectroscopically normal SN\,Ia 2003hv at $\sim 1$ year after explosion can be
better reproduced if the emitting mass, which is located mostly at $v <
10000$\,\kms, is significantly reduced with respect to what is predicted by a
standard explosion model such as W7. Delayed detonation models have a similar
density distribution as W7 in the inner layers 
\citep[][Fig. 3]{Hoflich95,Iwamoto99}.
The reduced mass in our model is required by the relatively high ionization of
the gas, as signalled by the unusual strength of the [\FeIII] optical emission
lines relative to the [\FeII] lines as well as by the unexpectedly low near-IR
flux, which again consists mostly of [\FeII] lines. 

\citet[][Fig. 8]{Leloudas09} have shown that the contribution of the near-IR
flux to the emitted spectrum increases from 5\,\% of the UVOIR flux \citep[][the
mid-IR flux is not included in this estimate]{Leloudas09} at $\sim 100$--200
days to $\sim 10$\,\% at about 1 year, and reaches a maximum of $\sim 40$\,\% at
$\sim 600$ days. We have tested whether our model can reproduce this trend. We
have computed spectra at different epochs and integrated the flux to determine
the relative near-IR contribution as a function of time. The results are show in
Figure 9: although our model with a modified inner density does not reproduce
the observed values exactly, it is much closer to the observed trend than the
W7-based model. This adds support to the idea that SN\,2003hv had a mass at low
velocities significantly smaller than what Chandrasekhar-mass models predict
\citep{Maeda10a}. 

We have not extended the calculations to beyond 600 days, because at those
epochs emission is expected to shift to the mid-IR, and many of the relevant
lines are not included in our code.

\begin{figure}
\includegraphics[width=90mm]{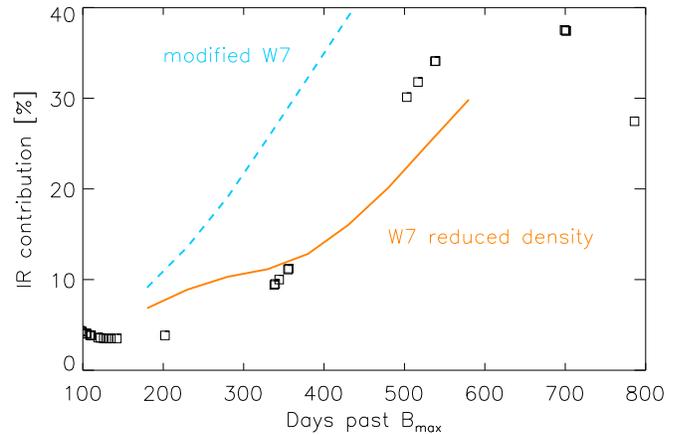}
\caption{The near-IR contribution to the UVOIR flux as a function of time:
the points are the measured fractions \citep[from][]{Leloudas09}, while the
orange line is the result of our model with reduces mass. 
We also show the expected behaviour of the modified W7 model (with a reduced
\Nifs\ mass) which we use to model the spectrum of SN\,2003hv. }
\label{near-IR}
\end{figure}

\section{Discussion: mass and risetime}

Our best-fit model has a mass of $M_\mathrm{ej} = 0.5 \Msun$ inside 10000\,\kms.
The \Nifs\ mass is $\approx 0.18 \Msun$.  By contrast, a model like W7 has a
mass of $0.9 \Msun$ inside the same velocity. Models based on W7, with varying
abundance distributions (in particular different \Nifs\ masses) have
successfully been used with our code to reproduce the nebular spectra of
SNe\,Ia of different luminosities \citep[\eg][]{Stehle2005,Mazzali08,Tanaka11}.
The difference we see in this case  suggests that SN\,2003hv may have been the
result of a different type of explosion, with a mass that is distributed
differently in velocity  and may be different from the Chandrasekhar mass.

\subsection{Mass estimates and uncertainties}

Here we address some uncertainties inherent to our mass estimates. The nebular
model presented above has a smaller mass than W7 in the inner part ($v <
10000$\,\kms). It preserves the density-velocity structure of W7 at higher
velocities, but those layers are not probed by nebular spectroscopy.  Our model
probably somewhat underestimates both the \Nifs\ mass and the mass at low
velocities, for two reasons. First, the low-level flux between the strong
optical emission lines is not well reproduced: our synthetic flux in the optical
region is lower than the observed one by $\sim 20$\%. Secondly, the flux
emerging in the mid-IR contributes $\approx(34 \pm 17)$\,\% to the total SN
luminosity 358 days after the explosion \citep{Leloudas09}, while this fraction
is only $\sim$ 3\% in our synthetic spectrum. 

The missing continuum in our synthetic model is the consequence of the absence
of UV radiation transport in our nebular code. It is known that UV excitation
and fluorescence are reponsible for such continua \citep{LiMcCray96}.  The
missing mid-IR flux is probably caused by atomic transitions at energies of
about 0.1 eV and less, which are responsible for mid-IR emission but are 
partially missing in the nebular code.  Since an increase in flux can be
obtained by increasing either the mass of \Nifs\ or the total emitting mass (in
both cases the energy deposited in the nebula increases), both the \Nifs\ mass
and the emitting mass (within 10000\,\kms) could be increased by a factor $1.25
\pm 0.10$. 
Since our model has an emitting mass of $0.5 \Msun$ within 10000\,\kms, when 
this correction is applied we get an emitting mass of $0.63\pm0.05 \Msun$. 
As for the \Nifs\ mass, in our models we obtained $0.18 \Msun$. This then
becomes $0.22\pm0.02 \Msun$. Both values are still significantly smaller than
the corresponding values in W7, which has a mass of $0.9 \Msun$ inside of
10000\,\kms\ and a \Nifs\ mass of $\sim 0.63\Msun$.  

Other sources of uncertainty can affect our result. 
The distance we adopted ($\mu =31.58$\,mag) was derived matching the $BVRI$
light curve of SN\,2003hv to template light curves \citep{Leloudas09}. This is
0.2\,mag larger than the corrected SBF distance to  the host galaxy of
SN\,2003hv, NGC\,1201 \citep[31.37\,mag,][]{Tonry01,Jensen03}. If we had used 
the SBF distance, both the \Nifs\ mass and the total mass  would be smaller 
than what we have derived by $\sim 10$\% (alternatively, either the \Nifs\ mass 
or the total mass might be smaller by up to $\sim 20$\,\%).  
The extinction to SN\,2003hv is quite small, so that our results are basically
insensitive to its exact value. 




We have tested whether the mass which is missing in the inner layers may be
located at high velocities by computing a model where we increased the density
of shells above 10000\,\kms\ so as to preserve the Chandrasekhar mass of the
ejecta. The corresponding spectrum, which is shown in Figure 4, is not easy to
distinguish from that of the model with a lower mass. This is because, even with
an increased mass, the density in the outer layers is still too low to trap
$\gamma$ rays and produce line emission at the epoch of the nebular spectrum of
SN\,2003hv. 

An abundance tomography experiment such as that performed for SN\,2005bl
\citep{Hachinger09} would be required to determine the mass and energy in the
outer layers. We plan to perform such an analysis in the near future.

\subsection{\Nifs\ mass and risetime}

\citet{Leloudas09} derive a \Nifs\ mass of $\sim 0.4 \Msun$ from the light
curve peak applying Arnett's rule \citep{Arnett82} as calibrated by
\citet{Stritzinger06} using the formula 
\begin{equation}
L = 2 \times 10^{43}~{\rm M}(^{56}{\rm Ni})/{\rm M}_{\odot} ~~~~~~ [{\rm erg~s}^{-1}],
\end{equation}
which is calibrated for a risetime $t_r = 19$\:days. 
They perform a simple light curve analysis and find that this value is however 
inconsistent with the late part of the light curve, which requires a \Nifs\ mass
of $\sim 0.2 \Msun$. They also show a synthetic nebular spectrum which is much
brighter than the observations. 

\citet{Leloudas09} suggest various possibilities to explain this discrepancy,
including an early near-IR catastrophe which causes the flux to shift to the
far-IR.  An early IR catastrophe seems not to be supported by the data, because
we have seen that the gas is still quite highly ionised even at later times, and
is charachterised by temperatures well in excess of 1000\,K. An IRC occurring
only in the innermost layers, a possibility suggested by \citet{Leloudas09},
would probably not transfer enough energy to the far-IR.

On the other hand, our model with a reduced mass at low velocity is capable of
reproducing the late-time flux for a \Nifs\ mass similar to what
\citet{Leloudas09} find from the late light curve. This leaves open the question
of why such different \Nifs\ masses can be derived from the early and the late
data, which usually are in good agreement \citep{Stritzinger06}. Although
Arnett's rule may not be highly accurate, a factor of 2 difference is a large
discrepancy. 

We suggest that a simple solution to the problem can be found if the density in
the inner part of the ejecta is reduced. The \Nifs\ mass was estimated from the
peak of the light curve using the formula above, which assumes that light curve
maximum occurs 19 days after the explosion \citep{Stritzinger06}. However, if
the light curve risetime of SN\,2003hv was significantly smaller than 19 days,
this discrepancy may be resolved. Using the dependence of the luminosity on the
\Nifs\ mass and the risetime as given in Eq.5 of \citet{StritzLeib05} yields the
curve shown in Figure 10. Adopting the estimate of the \Nifs\ mass from the
nebular models ($0.22 \Msun$) instead of the value obtained by
\citet{Leloudas09} from the peak of the light curve and an estimated risetime of
19 days ($0.4 \Msun$), \ie reducing the \Nifs\ mass by (0.22/0.40) = 0.55, a
risetime $t_r \approx 10$\:days would be inferred. This would reflect the rapid
decline of the light curve, while the \Nifs\ mass estimate from the nebular
spectrum would hardly be affected by this choice of risetime, since the decay
time of \Cofs, which powers the SN light curve at these advanced stages, is much
longer. 

\begin{figure}
\includegraphics[width=90mm]{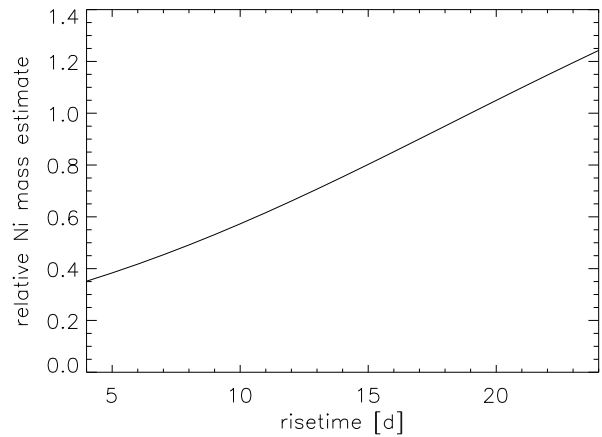}
\caption{Effect on the risetime of a modified estimate of the \Nifs\ mass,
including both \Nifs\ and \Cofs\ decay. }
\label{rise}
\end{figure}

A risetime of $\sim 10$ days is short for SNe\,Ia. Estimates of the typical
risetimes in the literature have decreased from $\approx 19.5$ days, a result
based on extrapolating the rising part of the light curve
\citep{Riess99,Aldering2000}, to values between $\approx 19.1$ days
\citep{Conley06} and $\approx 17.5$ days \citep{Garg07,Hayden10}, based on
direct observations of SNe. Both of these values are significantly larger than
what would be required in order to bring the early- and late-time estimates of
the \Nifs\ mass in SN\,2003hv into agreement.  On the other hand, methods based
on fitting the rising part of the light curve yield shorter risetimes, which are
possibly correlated to the decline rate \citep{Strovink07}, when only SNe with a
sufficiently well sampled pre-maximum photometry are analysed. The fastest
decliner in \citet{Hayden10} and \citet{Strovink07} is SN\,2004eo 
\citep[\Deltam\,$ = 1.47$\,mag;][]{Pasto07}, for which they estimate  $t_r
\simeq 16.6$\:days. However, shorter risetimes have been inferred or observed.
SN\,1994D  (\Deltam\,$ = 1.34$\,mag) is quoted to have $t_r \simeq 15.4$\:days 
\citep{Strovink07,Hayden10}. \citet{Hayden10} observed a few SNe with $t_r \sim
13 - 15$ days. The incidence of rapid risers is much higher for dimmer SNe
(\Deltam\,$ \gsim 1.5$\,mag), although these SNe are much more seldom seen than
brighter ones.  Extrapolating the relation in Figure 2 of \citet{Strovink07} to
the decline rate of SN\,2003hv, \Deltam\,$ = 1.61$\,mag, a risetime of $\sim
13-15$ days is expected, in agreement with the result of \citet{Hayden10}.  

A further suggestion of a short risetime of SN\,2003hv comes from an attempt to
compute the light curve properties in an approximate way as done in
\citet{Zorro}. Adopting a total mass of $1.12 \Msun$, and assuming that this is
distributed for $0.22 \Msun$ as \Nifs, $0.07 \Msun$ as stable Fe-group material,
and that most of the remaining mass located inside of $11000$\,\kms ($0.4
\Msun$) has been burned to IME, this yields a nuclear energy release of $\approx
10^{51}$\,erg. Subtracting from this the binding energy of a white dwarf of
$\sim 1.12 \Msun$, $\sim 2\cdot 10^{50}$\,erg, an explosion energy \KE\ $\approx
8\cdot 10^{50}$\,erg is obtained. Combined with an estimate of the opacity as
discussed in \citet{Zorro}, this yields a characteristic bolometric LC width of
$\sim 15$ days. Looking at Fig. 2 of \citet{Zorro}, this would correspond to a
peak luminosity of only $\sim 3\cdot 10^{42}$\,erg\,s$^{-1}$, which is much less
than the observed value $\sim 10^{43}$\,erg\,s$^{-1}$. Therefore SN\,2003hv does
not seem to comply with the correlations established for SNe\,Ia: the \Nifs\
mass derived assuming a relation to the peak luminosity calibrated for a
risetime of 19 days is too high for the decline rate of the SN.

\begin{figure}
\includegraphics[width=90mm]{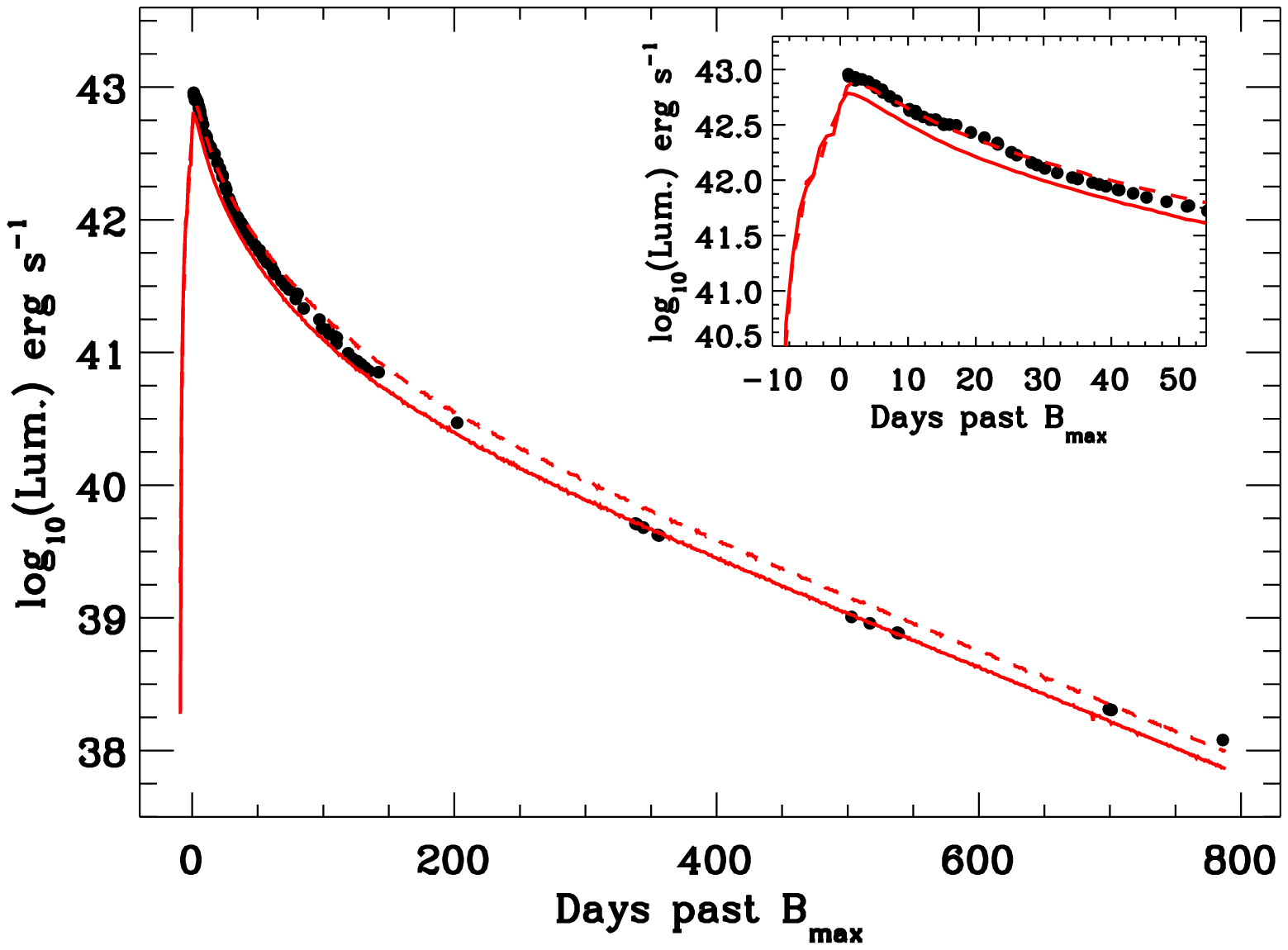}
\caption{Two synthetic bolometric light curves compared to the UVOIR light curve
of SN\,2003hv \citep[dots,][]{Leloudas09}. The fully drawn line is a model 
computed for the density structure and \Nifs\ mass derived from the nebular
spectrum models. It has M$ = 1.0 \Msun$, M(\Nifs)$ = 0.18 \Msun$, and \KE$ =
1.24\: 10^{51}$\,erg. The dashed line is a model where both the \Nifs\ mass and 
the total mass located at velocities $v < 8000$\,\kms\ are multiplied by a 
factor 1.25, reflecting inaccuracies of the nebular-epoch modelling (see 
Section 5.1). This model has M$ = 1.12 \Msun$, M(\Nifs)$ = 0.22 \Msun$, and 
\KE$ = 1.26\: 10^{51}$\,erg. Both models have risetimes of $\sim 14$ days.}
\label{rise}
\end{figure}

We finally tested the possible risetime for SN\,2003hv using the same grey
Montecarlo light curve code which could successfully reproduced the bolometric
light curves of other SNe\,Ia \citep[\eg][]{Stehle2005,Mazzali08,Tanaka11}. We
use the ``reduced density" model derived from nebular spectroscopy and shown in
Fig. 6. The model has a mass of $0.5 \Msun$ inside 10000\,\kms. Outside of this
velocity we simply assume the W7 density distribution. The total mass is
therefore $1.12 \Msun$, the \Nifs\ mass is $0.22 \Msun$, and the kinetic energy
\KE$ = 1.26\cdot 10^{51}$\,erg.  We also used the same definition of the opacity
as a function of chemical species \citep{Zorro} as in those papers. The
resulting bolometric light curve is shown in Figure 11. The risetime of the
synthetic light curve is short, $\sim 13.5$ days.  We compare this synthetic
light curve to the UVOIR light curve of SN\,2003hv \citep{Leloudas09} in Fig.11,
and find good agreement in shape from the earliest observations, which are near
maximum, to the latest ones, $\sim 800$ days later.  The rising part of the
light curve is not constrained by the data. Since we could derive the
distribution of \Nifs\ from the nebular spectra only, we have not included any
\Nifs\ at $v > 10000$\,\kms. If some \Nifs\ was present at higher velocities,
the rising part of the light curve would be smoother, and the peak may be
reached a little earlier. Figure 11 also shows a model where both the \Nifs\
mass and the total mass located at velocities $v < 8000$\,\kms\ are multiplied
by a factor 1.25, reflecting inaccuracies of the nebular-epoch modelling (see
Section 5.1). The light curve for this model reaches peak somewhat later ($\sim
14$ days after explosion) and is overall $\sim 0.2$ mag brighter. In both cases
the inconsistency between the peak and the tail of the light curve is
eliminated.

The reason for the rapid rise is that not only the inner ejecta have low
density, and that the total mass of Fe-group species is quite small ($0.24
\Msun$), which keeps the opacity low. This is compounded by a normal ejecta
velocity (the maximum-light spectrum of SN\,2003hv has a \SiII\ line velocity of
$\approx 10500$\,\kms). 
Although what we have used here is by no means a realistic explosion model, the
result that we obtained is nevertheless suggestive.

\section{Conclusions}

The high ionization and low near-IR flux in the nebular phase of SN\,2003hv can
be best explained if this SN\,Ia ejected less mass at low velocities than in
standard single-degenerate, Chandrasekhar-mass explosion models.  In particular,
the mass of stable NSE elements is greatly reduced. The smaller mass may lead to
a short rise time, which would also solve the apparent inconsistency between the
\Nifs\ mass derived from the peak of the light curve and the nebular epoch.
Therefore, the \Nifs\ mass obtained from the peak luminosity using a relation
which has been calibrated for a rise time of 19 days may be incorrect.

Two alternative scenarios exist to the single-degenerate, Chandrasekhar-mass
one. One is the explosion of sub-Chandrasekhar mass white dwarfs. This involves
a CO white dwarf accreting He from a companion and an edge-lit explosion
\citep{Livne90,WoosleyWeaver94,LivneArnett95,Fink10}. \citet{Sim10} calculated
light curves of sub-Chandrasekhar models which do not consider how the 
explosion is triggered. Their more massive models predict risetimes of $\sim 18$
days. In those models, however, Fe-group elements are much more abundant than in
the model we have derived, which is likely to result in a larger optical
opacity. Their least massive model ($0.88 \Msun$) has a risetime of 14 days,
which is comparable to our result. In this model \Nifs\ is more concentrated
towards the centre than the distribution we have derived for SN\,2003hv
\citep[Fig.7;][Fig. 1]{Sim10}, and the kinetic energy is smaller ($8.6 \cdot
10^{50}$\,erg compared to $\sim 1.25 \cdot 10^{51}$\,erg).  The diffusion time
of optical photons may therefore be comparable in SN\,2003hv despite the
slightly larger mass. 

The other scenario is the merging of two white dwarfs
\citep{IbenTutukov84,Webbink84}, and in particular the case of dynamical mergers
\citep{Pakmor10}. This results in an explosion ejecting at least a Chandrasekhar
mass, but the two merging white dwarfs had mass smaller than the Chandrasekhar
mass, and consequently lower central densities. This is reflected in the density
distribution of the SN ejecta. 

In Fig. 12 we compare the density structures of W7, of a merger model with total
mass $1.8 \Msun$ \citep{Pakmor10}, of two sub-Chandrasekhar models, with total
mass 0.88 and $1.06 \Msun$ and \Nifs\ mass of 0.07 and $0.43 \Msun$,
respectively \citep{Sim10}, and the density structure we derived from our
nebular model. We stress that our model is reliable only inside of 10000\,\kms.
In the inner ejecta ($v < 10000$\,\kms) the two sub-Chandrasekhar models and the
density distribution we derived actually look quite similar. A model with \Nifs\
mass similar to that of SN\,2003hv is not available. 

\begin{figure}
\includegraphics[angle=-90,width=90mm]{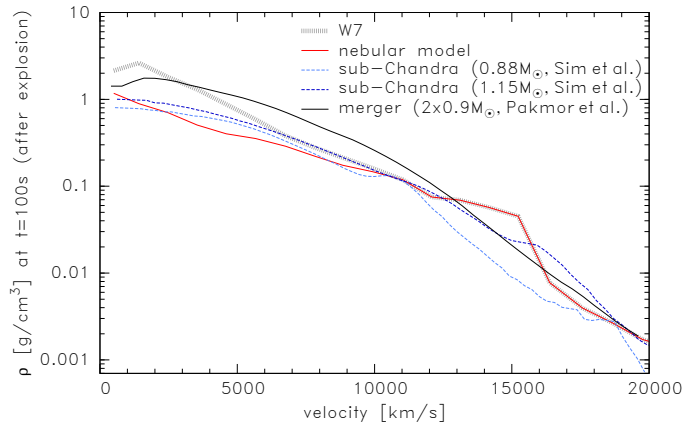}
\caption{A comparison of various density profiles:
the W7 density distribution (grey);
the modified model with a reduced density in the inner layers obtained from our
nebular calculations (red);
a model of an explosion following the dynamical merging of two white dwarfs
\citep{Pakmor10} (black), 
two different sub-Chandrasekhar models (blue) \citep{Sim10}. }
\label{density}
\end{figure}

It will be interesting to submit all non-standard models to the test of nebular
spectroscopy. Clearly, the availability of flux-calibrated near-IR spectra at
late times is an important piece of information if we want to discriminate among
different models.  

The exact incidence of SNe with a non-standard origin is not known. In most
SNe\,Ia, optical nebular spectra do not suggest a high \FeIII/\FeII\ ratio. 
A dominance of \FeIII\ over \FeII\ is deduced also in the innermost core of
SN\,1991bg, for which an improved fit to a series of nebular spectra can also be
obtained by slightly reducing the mass at low velocities (Mazzali et al., in
preparation). Both SNe 2003hv and 1991bg are characterised by a low luminosity,
although the former is still spectroscopically normal at maximum. A possibility
is that all SNe\,Ia at the dim end of the distribution are the result of
non-standard explosions, possibly of sub-Chandrasekhar mass white dwarfs.

The final question is whether the unexpectedly low near-IR flux, which is one of
the characteristics that leads to a low mass estimate for SN\,2003hv, is typical
of other SNe\,Ia. Late-time near-IR spectroscopy is available only for two other
SNe\,Ia, 2003du and 2005W \citep{Motohara06}. There are also a couple of
photometric observations in the near-IR, for SNe\,2000cx and 2001el
\citep{Sollerman04,Stritzinger07}. In both of these cases, the ratio of near-IR
to optical flux indicated a lower near-IR flux than expected from standard
models like W7. While it is possible that other SNe may be accommodated in 
scenarios where the inner mass is reduced with respect to classical models based
on this evidence, we refrain from making that suggestion here. Near-IR and
mid-IR spectrophotometry would be required to be more confident about any such
statement. In particular, SN\,2000cx was very bright, indicating that most of
the inner part of the white dwarf progenitor had been burned to NSE. Also,
spectroscopic modelling did not indicate the need for a low density in the
innermost layers \citep{Zorro}.

In conclusion, near-IR spectrophotometry of SNe\,Ia in the nebular phase has
proved to be a very important source of information in the case of SN\,2003hv.
It is to be hoped that such observations will be performed routinely for nearby
SNe\,Ia (as well as other Type I SNe) in the future.


\section*{Acknowledgments} We thank Fritz R\"opke and Rudiger Pakmor for
providing the density structure of their explosion models. 
We gratefully acknowledge helpful conversations with
Wolfgang Hillebrandt, and thank Keiichi Maeda for useful
comments on an earlier version of this manuscript.
P.A.M. and S.B. are partially supported by PRIN-INAF 2009 project 
``Supernovae Variety and Nucleosynthesis Yields".
S.T. acknowledges support by the TRR33 ``The Dark Universe" of the German
Research Foundation .


\bsp

\label{lastpage}

\end{document}